\documentclass[12pt,a4paper]{article}

\usepackage{circuitikz}
\usepackage{tikz}
\usepackage{dashrule}

\usepackage{authblk}
\usepackage{titlesec}
\usepackage{aas_macros}
\usepackage{multicol}
\usepackage[hang,flushmargin]{footmisc}

\usepackage{amsmath}
\usepackage{amssymb}
\usepackage{amsthm}
\usepackage{appendix}
\usepackage{braket}
\usepackage[english]{babel}
\usepackage{caption}
\usepackage[sectionbib]{chapterbib}
\usepackage{xcolor}
\usepackage{enumitem}
\usepackage{epsfig}
\usepackage{epstopdf}
\usepackage{fancybox}
\usepackage{fancyhdr}		
\usepackage{fancyvrb}
\usepackage[Bjarne]{fncychap}
\usepackage[inner=2cm,outer=2cm,top=1cm,bottom=1cm,includehead,includefoot,heightrounded]{geometry}
\usepackage{graphicx}
\usepackage{latexsym}%
\usepackage{longtable}
\usepackage{lscape}%
\usepackage{mathrsfs}
\usepackage{mathtools}
\usepackage{microtype}%
\usepackage{multirow}
\usepackage[authoryear,round]{natbib}
\usepackage[parfill]{parskip}
\usepackage{pdfsync}
\usepackage{psfrag}
\usepackage[crop=pdfcrop,process=auto]{pstool}
\usepackage[dvips]{rotating}   	
\usepackage{siunitx}
\usepackage{subcaption}
\usepackage{tensor}
\usepackage{tikz}
\usepackage{ulem}
\usepackage{upquote}%
\usepackage{wasysym}%
\usepackage{wrapfig}
\usepackage[colorlinks=true, pdfstartview=FitV, linkcolor=blue, citecolor=blue, urlcolor=blue]{hyperref}

\titleformat*{\section}{\large\bfseries}

\title{\vspace{-72pt}Throwing $\pi$ at a wall}
\author[1,$\dagger$]{M. Z. Rafat}
\author[1,2,$\dagger$]{D. Dobie}
\affil[1]{Sydney Institute for Astronomy, School of Physics, University of Sydney, NSW 2006, Australia}
\affil[2]{ATNF, CSIRO Astronomy and Space Science, PO Box 76, Epping, NSW 1710, Australia}
\affil[$\dagger$]{These authors contributed equally to this work}
\date{}

\begin{document}
\maketitle

\begin{abstract}
We discuss a method for calculating $ \pi $ using elastic collision between two masses $ M $ and $ m $ with $ X = m/M = 10^{2(1-d)} $ where $ d $ is an integer, and a wall. The total number of collisions between $ M $, $ m $ and the wall corresponds to the first $ d $ digits of $ \pi $.
\end{abstract}
\vspace{-36pt}

\section*{Introduction}
\vspace{-12pt}
Obtaining an accurate value of $\pi$ is one of the oldest problems in mathematics. Early approximations of $\pi$ accurate to one decimal place dating back nearly 4000 years have been discovered in Babylon and Egypt \citep{beckmann1971history}. The oldest known algorithm for calculating $\pi$ is attributable to Archimedes who placed upper and lower bounds on its value by inscribing and circumscribing n-sided regular polygons on a circle in his treatise \textit{Measurement of a Circle}. The advent of computers has enabled much faster and more accurate calculations of $\pi$, and the current best algorithm is capable of computing it to 22 trillion decimal places\footnote{\url{https://pi2e.ch/blog/2016/10/31/hexadecimal-digits-of-pi/}}. 
A recent video published by \textit{3Blue1Brown}\footnote{\url{https://www.youtube.com/watch?v=HEfHFsfGXjs}} brought our attention to an unusual method of calculating $\pi$ originally discovered by \citet{galperin2003}. While the efficiency of this method pales in comparison to modern algorithms, it is certainly a novel approach.

\begin{figure*}[!t]
    \centering
    \begin{tikzpicture}
    \draw[thick] (0,0) -- (12,0);
    \draw[thick] (0,0) -- (0,3.5);
    
    \draw[thick,fill=lightgray] (4,0) rectangle (4+1.5,1.5);
    \node at (4.75,0.75) {$m$};
    
    \draw[thick,fill=lightgray] (8,0) rectangle (8+2,2);
    \node at (9,1) {$M$};
    
    \draw[thick,->] (8,1) -- (7,1) node[anchor=north west]{$~\,v_0$};
	\end{tikzpicture}
	
	\begin{tikzpicture}
    \draw[thick] (0,0) -- (12,0);
    \draw[thick] (0,0) -- (0,3.5);
    \draw[thick,color=white] (0,3.5) -- (0,4);
    
    \draw[thick,fill=lightgray] (1.2,0) rectangle (1.2+1.5,1.5);
    \node at (1.2+0.75,0.75) {$m$};
    \draw[thick,->] (1.2,0.75) -- (0.2,0.75) node[anchor=north west]{$~\,u_1$};
    
    \draw[thick,fill=lightgray] (4.5,0) rectangle (4.5+2,2);
    \node at (4.5+1,1) {$M$};
    \draw[thick,->] (4.5,1) -- (4.5-1,1) node[anchor=north west]{$~\,v_1$};
    
	\end{tikzpicture}
	
	\begin{tikzpicture}
    \draw[thick] (0,0) -- (12,0);
    \draw[thick] (0,0) -- (0,3.5);
    \draw[thick,color=white] (0,3.5) -- (0,4);
    
    \draw[thick,fill=lightgray] (0.3,0) rectangle (0.3+1.5,1.5);
    \node at (0.3+0.75,0.75) {$m$};
    \draw[thick,->] (0.3+1.5,0.75) -- (0.3+1.5+1,0.75) node[anchor=north east]{$u_1~\,$};
    
    \draw[thick,fill=lightgray] (3.8,0) rectangle (3.8+2,2);
    \node at (3.8+1,1) {$M$};
    \draw[thick,->] (3.8,1) -- (3.8-1,1) node[anchor=north west]{$~\,v_1$};
    
	\end{tikzpicture}

	\caption{The first two collisions of the process. Top: initial positions and velocity. Middle: after the collision of both masses. Bottom: after the second mass has reflected from the wall.}
	\label{fig:collisions}
\end{figure*}
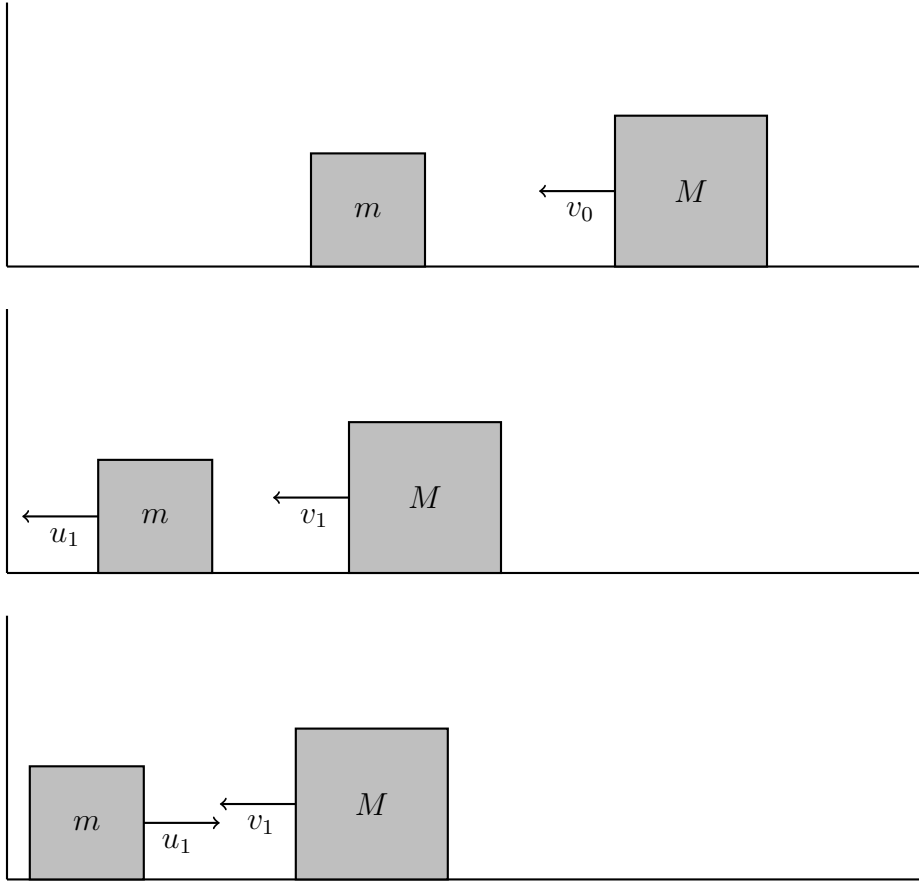

An object of mass $ M $ moves along a frictionless surface towards a second object of mass $m$ which is stationary, before colliding perfectly elastically with it (i.e. no energy is lost in the collision), propelling the second object towards an immovable wall. The second object undergoes perfectly elastic collision with the wall and is reflected back off the wall and collides again with the first mass as seen in the diagram in Figure \ref{fig:collisions}. This process repeats until both objects are moving away from the wall and the speed of the first object exceeds the speed of the second. \citet{galperin2003} found that if the ratio of masses, $ X = m/M$, is of the form $10^{2(1-d)}$, where $ d $ is a positive integer, then the number of collisions is an integer consisting of the first $d$ digits of $\pi$.

\citet{galperin2003} and \citet{aretxabaleta2017} consider the position and velocity of both objects and treat the problem as a \textit{billiard system} where particles collide with each other and immovable boundaries. In this paper we take a different approach to achieve the same result and instead consider the evolution of the velocity phase-space of the system.

\section*{Mathematical formulation}
\vspace{-12pt}
Our system consists of masses $ M $ and $ m $ with velocities $ {\bf v} $ and $ {\bf u} $, respectively, with magnitudes $ v $ and $ u $. Collisions between the masses conserve both energy and momentum as we assume perfectly elastic collisions and ignore all external forces. The collision between mass $ m $ and the wall switches the sign of the velocity of $ m $. Therefore each collision with the wall results in change of momentum of the system while leaving the kinetic energy unaffected. The kinetic energy and momentum of the system may be written, respectively, as
\begin{equation}\label{eq:KP}
    \frac{1}{2}M v^2 + \frac{1}{2}m u^2 = K = \frac{1}{2}M v_0^2,
    \quad\text{and}\quad
    M {\bf v} + m {\bf u} = {\bf P},
\end{equation}
where $ v_0 $ is the initial speed of $ M $. We choose the positive direction to be along the negative horizontal axis (towards the wall). Henceforth we denote $ {\bf v} $ and $ {\bf u} $ simply as $ v $ and $ u $ with the direction indicated by their signs. The velocity of $ M $ and $ m $ after they collide is given by
\begin{equation}\label{eq:v}
    v = \frac{(1 - X) v' + 2X u'}{1 + X},
    \quad\text{and}\quad
    u = \frac{2v' + (X - 1) u'}{1 + X},
\end{equation}
where $ X = m/M $ and $ v' $ and $ u' $ denote the velocity of $ M $ and $ m $ before the collision, respectively. We may express~\eqref{eq:KP} as
\begin{equation}\label{eq:KP2}
    \frac{v^2}{a^2} + \frac{u^2}{b^2} = 1,
    \quad\text{and}\quad
    v + X u = \frac{P}{m}
\end{equation}
where $ a = v_0 $ and $ b = v_0/\sqrt{X} $. Valid solutions for $ v $ and $ u $ are then given by the points of intersection of a straight line arising from the momentum equation with the kinetic energy ellipse. We may express the collision of $ M $ and $ m $ using~\eqref{eq:v} as $ {\bf w} = {\bf A} {\bf w}'/(1 + X) $ where
\begin{equation}
    {\bf w} =
    \begin{pmatrix}
    v\\[1ex]
    u
    \end{pmatrix},
    \quad
    {\bf A} =
    \begin{pmatrix}
    1-X & 2X\\[1ex]
    2 & X-1
    \end{pmatrix},
    \quad
    {\bf w}' = 
    \begin{pmatrix}
    v'\\[1ex]
    u'
    \end{pmatrix}.
\end{equation}
Similarly the collision with the wall may be expressed as $ {\bf w} = {\bf W} {\bf w}' $ with $ {\bf W} = {\rm diag}(1,-1) $. Denoting the velocity after $ n^{\rm th} $ collision with a subscript $ n $ allows us to write for $ n $ odd and even, respectively, as
\begin{figure*}[!b]
    \centering
    \includegraphics[width=0.8\linewidth]{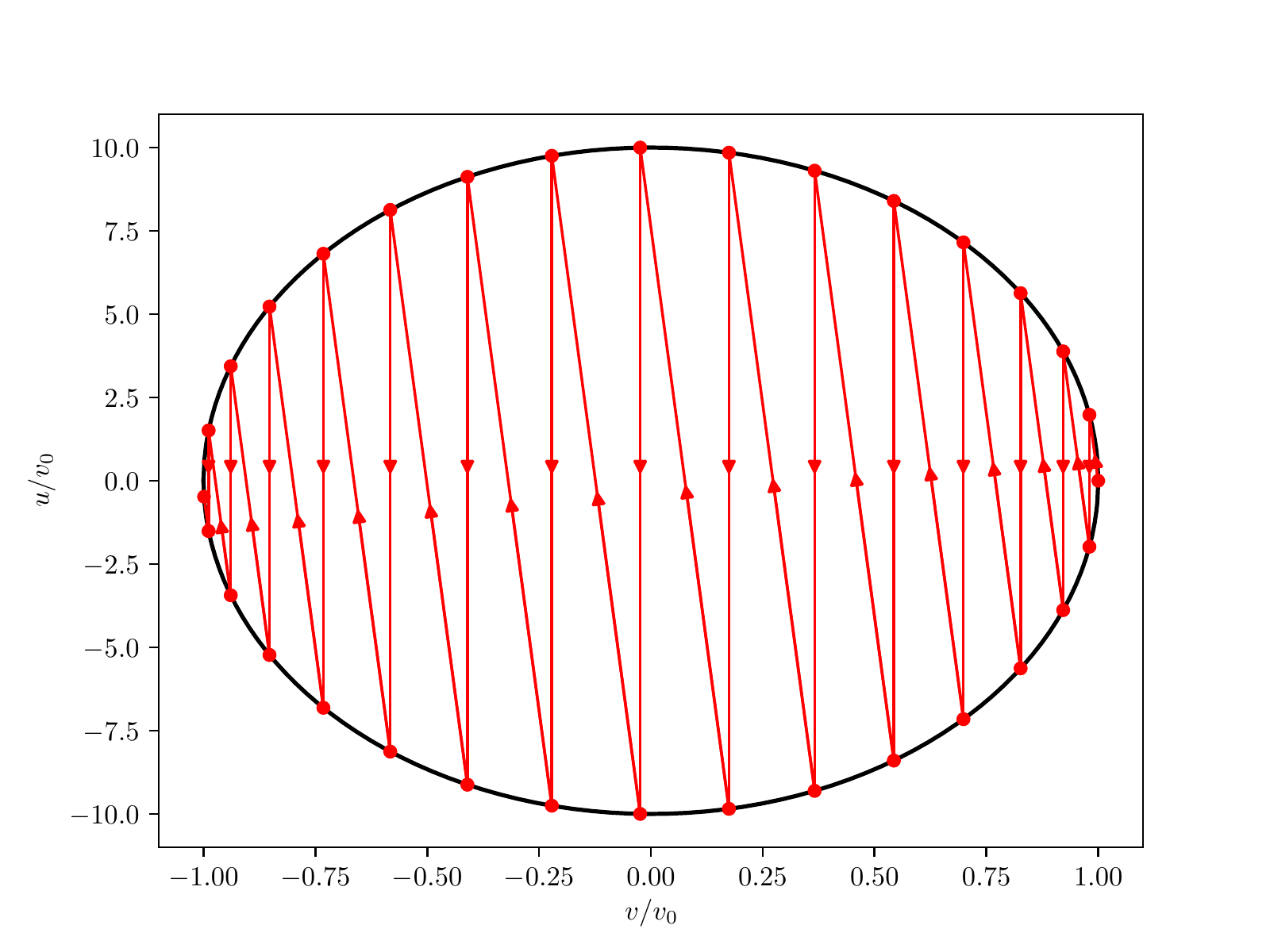}
    \caption{Path of the system (red) as it traverses the phase space for a mass ratio $X=10^{-2} $, giving 31 collisions. The ellipse given by conservation of energy constraints is shown in black, with the position of the velocity vector at each stage denoted by red circles.}
    \label{fig:particle_paths}
\end{figure*}
\begin{equation}\label{eq:wn}
    {\bf w}_{n} = \frac{{\bf H}_{\rm o}^{(n-1)/2}{\bf w}_1}{(1+X)^{(n-1)/2}},
    \quad\text{and}\quad
    {\bf w}_{n} = \frac{{\bf H}_{\rm e}^{n/2}{\bf w}_0}{(1+X)^{n/2}},
\end{equation}
where
\begin{equation}
    {\bf H}_{\rm o} = {\bf A}{\bf W} 
    =
    \begin{pmatrix}
    1-X & -2X\\[1ex]
    2 & 1-X
    \end{pmatrix},
    \quad
    {\bf H}_{\rm e} = {\bf W}{\bf A} 
    =
    \begin{pmatrix}
    1-X & 2X\\[1ex]
    -2 & 1-X
    \end{pmatrix},
    \quad\text{and}\quad
    {\bf w}_1 = \frac{{\bf A} {\bf w}_0}{1 + X}.
\end{equation}
The odd values of $ n $ corresponds to collision between $ M $ and $ m $ while the even values correspond to collision between $ m $ and the wall. It follows from~\eqref{eq:wn} that for odd and even $ n $ we have
\begin{equation}\label{eq:wn2}
    {\bf w}_{n + 2} = \frac{{\bf H}_{\rm o}{\bf w}_n}{1+X},
    \quad\text{and}\quad
    {\bf w}_{n + 2} = \frac{{\bf H}_{\rm e}{\bf w}_n}{1+X},
\end{equation}
respectively. We may write $ {\bf H}_{\rm o} = {\bf P}_{+} {\bf D} {\bf P}_{+}^{-1}$ and $ {\bf H}_{\rm e} = {\bf P}_{-} {\bf D} {\bf P}_{-}^{-1}$ with
\begin{equation}
    {\bf P}_\pm =
    \begin{pmatrix}
        \mp i\sqrt{X} & \pm i\sqrt{X}\\[1ex]
        1 & 1
    \end{pmatrix},
    \quad
    {\bf P}_\pm^{-1} = \frac{1}{2i\sqrt{X}}
    \begin{pmatrix}
        \mp 1 & i\sqrt{X}\\[1ex]
        \pm 1 & i\sqrt{X}
        \end{pmatrix},
    \quad
    {\bf D} = (1-X) 
    \begin{pmatrix}
        \lambda_{-} & 0\\[1ex]
        0 & \lambda_{+}
    \end{pmatrix},
\end{equation}
where
\begin{equation}
    \lambda_\pm = 1 \pm i\frac{2\sqrt{X}}{1-X} = \frac{(1+X)e^{\pm i\theta_0}}{1-X},\quad
    \theta_0 = \tan^{-1}\left(\frac{2\sqrt{X}}{1-X}\right).
\end{equation}
This allows us to express~\eqref{eq:wn} as
\begin{equation}\label{eq:wnfinal1}
    {\bf w}_{n} = {\bf R}_+\left[(n-1)/2\right]{\bf w}_1,
    \quad\text{and}\quad
    {\bf w}_{n} = {\bf R}_-\left[n/2\right]{\bf w}_0,
\end{equation}
for $ n $ odd and even, respectively, where
\begin{equation}
    {\bf R}_\pm\left[n\right] = \frac{{\bf P}_\pm{\bf D}^{n}{\bf P}_\pm^{-1}}{(1+X)^n} = 
    \begin{pmatrix}
        \cos(n\theta_0) & \mp \sqrt{X}\sin(n\theta_0)\\[1ex]
        \frac{\pm 1}{\sqrt{X}}\sin(n\theta_0) & \cos(n\theta_0)
    \end{pmatrix}.
\end{equation}
The matrices $ {\bf R}_+[n] $ and $ {\bf R}_-[n] $ are rotation matrices which, respectively, rotate a vector $ {\bf w} $ anti-clockwise and clockwise along an ellipse of the form $ v^2/k + u^2/(k/X) = 1 $ for any real $ k $. This implies that for $ n $ odd (even) the velocity $ {\bf w}_n $ is obtained from $ {\bf w}_1 $ ($ {\bf w}_0) $ through an anti-clockwise (clockwise) rotation along an ellipse of the form $ v^2/k + u^2/(k/X) = 1 $ for any real $ k $. We can express~\eqref{eq:wnfinal1} as
\begin{equation}\label{eq:wnfinal2}
    v_n = v_0 \cos\left(s\theta_0\right),\quad
    u_n = (-1)^{n+1}\frac{v_0}{\sqrt{X}}\sin\left(s\theta_0\right),
\end{equation}
where $ s = (n+1)/2 $ for odd values of $ n $, and $ s = n $ for even values of $ n $. This of course denotes the same ellipse as given by~\eqref{eq:KP2}.

Figure~\ref{fig:particle_paths} shows the ellipse described by the kinetic energy equation~\eqref{eq:KP2} for $ X = m/M = 0.01 $ and $ d = 2 $. The dots lying on the ellipse denote points $ (v_n, u_n) $ as given by~\eqref{eq:wnfinal2} with the arrows trace their development in phase space. The process starts at $ (v_0, 0) $, corresponding to $ (1, 0) $, and continues until the speed of the larger mass, $ v_{n} $, exceeds the speed of the smaller mass, $ u_{n} $, and the larger mass is moving away from the wall: $ |v_{n}| > |u_{n}| $ and $ v_n < 0$. Geometrically this corresponds to ${\rm w}_{n}$ being in the third quadrant of the energy ellipse. 

Each solution $ {\bf w}_n = (v_n, u_n) $ as given by~\eqref{eq:wnfinal2} is of the form $ {\bf w}_n = (a\cos\theta_n, b\sin\theta_n) $ so that the angle $ {\bf w}_n $ makes with the horizontal axis satisfies $ \tan\theta_n = au_n/bv_n $. In particular for the first collision the angle is $ \theta_0 $ and the area subtended is given by
\begin{equation}
    A_0 = \frac{1}{2}ab\tan^{-1}\left(\frac{2\sqrt{X}}{1-X}\right)
        = \frac{1}{2}ab\theta_0.
\end{equation}
The area subtended by successive solutions $ {\bf w}_n = (v_n, u_n) $ and $ {\bf w}_{n+2} = (v_{n+2}, u_{n+2}) $ in the upper/lower portion of the ellipse is given by
\begin{equation}\label{eq:An}
    A_n = \frac{1}{2}ab\lvert\theta_{n+2} - \theta_n\rvert = \frac{1}{2}ab\bigg|\tan^{-1}\left(\frac{ab(v_n u_{n+2} - v_{n+2} u_n)}{b^2 v_n v_{n+2} + a^2 u_n u_{n+2}}\right)\bigg|,
\end{equation}
where the absolute value ensures that $ A_n $ is positive for both positive (odd $ n $) and negative (even $ n $) angles. Now, using~\eqref{eq:wn2} we have
\begin{equation}
    v_n = \frac{(1-X)v_n \mp 2Xu_n}{1+X},\quad
    u_n = \frac{\pm2v_n + (1-X)u_n}{1+X},
\end{equation}
where upper (lower) sign corresponds to odd (even) values of $ n $. This then gives
\begin{align}
    ab(v_nu_{n+2}-v_{n+2}u_n)
        & = \frac{\pm2v_0^2}{\sqrt{X}(1+X)}(v_n^2 + Xu_n^2),\\
    b^2v_nv_{n+2} + a^2 u_n u_{n+2}
        & = \frac{v_0^2}{X}\frac{(1-X)}{1+X}(v_n^2 + Xu_n^2).
\end{align}
Substituting into~\eqref{eq:An} gives
\begin{equation}
    A_n = \frac{1}{2}ab\tan^{-1}\left(\frac{2\sqrt{X}}{1-X}\right) \equiv A_0
\end{equation}
Therefore the area subtended by all adjacent solutions are equal.
\begin{figure}
    \centering
    \includegraphics[width=0.8\linewidth]{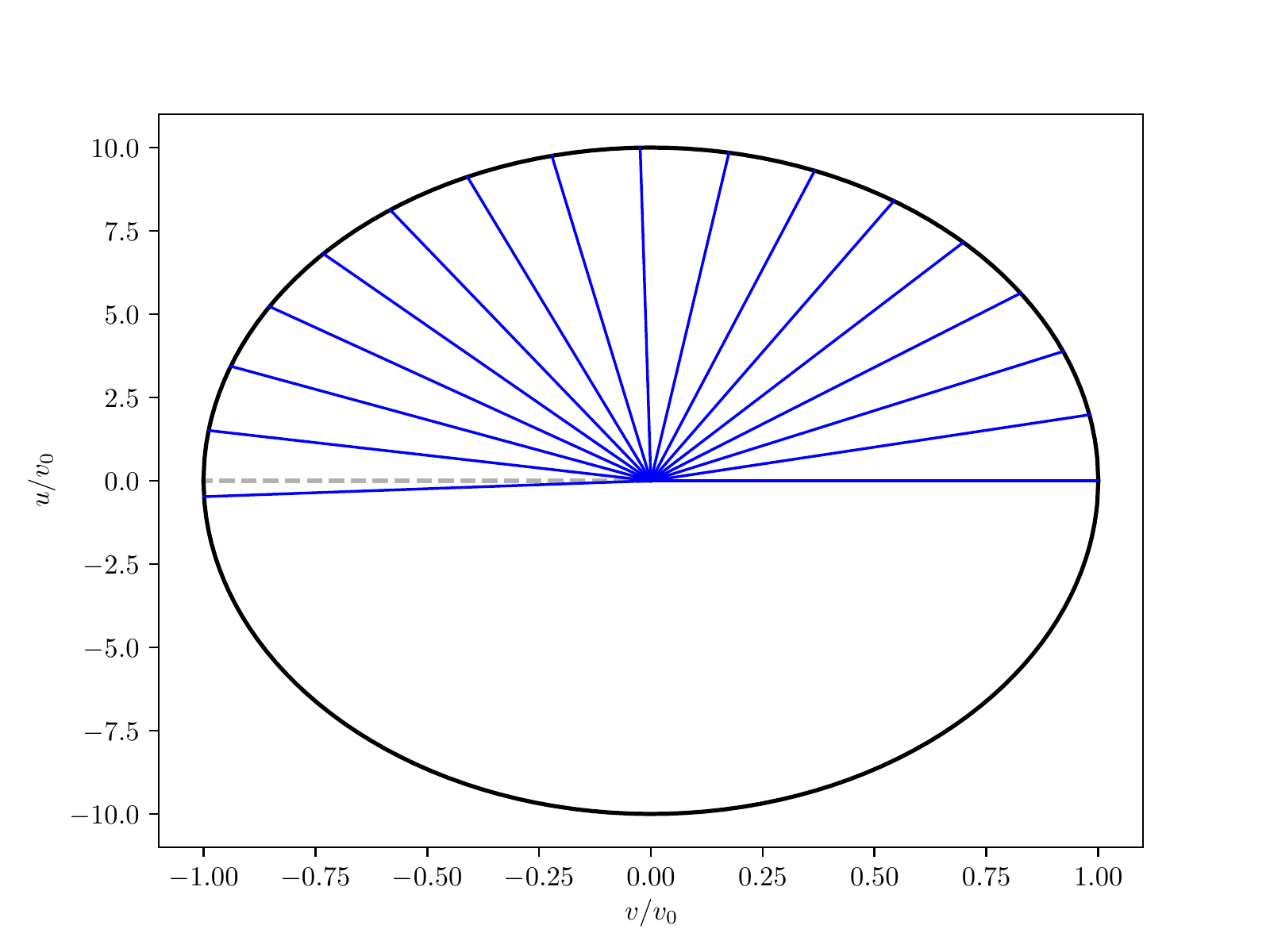}
    \caption{The energy ellipse (black) divided into equal segments (blue) by each stage of the process for $X=10^{-2}$. Only collisions between the objects are shown. The dashed grey line extends from the origin to $(-1, 0)$, showing that total area of the segments is slightly larger than the area of the semi-ellipse.}
    \label{fig:ellipse_segments}
\end{figure}
Now, let $ n $ denote the ratio of area of the energy ellipse and $ A_0 $:
\begin{equation}
    N = \frac{\pi ab}{A_0} = \frac{2\pi}{\tan^{-1}\left(\frac{2\sqrt{X}}{1-X}\right)}
    = \frac{\pi}{\sqrt{X}}\left(\sum_{k = 0}^\infty\frac{(-1)^k X^k}{2k + 1}\right)^{-1}.
\end{equation}
The series is absolutely convergent for $ |X| < 1 $. For $ X = 1 $ (equal mass) we have $ N = 4 $ and for $ X = 10^{2(1-d)} $, with $ d \geq 2 $, we may approximate $ N $ as
\begin{equation}
    \label{eq:def_N}
    N \approx \pi 10^{d-1},
\end{equation}
with maximum error of $ (\pi/3)10^{1-d} \ll 1 $ for $ d \geq 2 $. This number approximately corresponds to the number of sectors that the energy ellipse is divided into, $ N_{\rm total} $, noting that $ N_{\rm total} A_{0} > \pi ab $ for $ d \geq 2 $, as the final velocity vector $ \mathbf{w}_n $ is not precisely at $ (-v_0,0) $ as shown in Figure~\ref{fig:ellipse_segments}.

However, since the total number of collisions must be an integer, we can overcome this discrepancy by rounding, which gives
\begin{equation}
    \label{eq:def_N_tot}
    N_{\rm total} = {\rm ceil}(N) - 1.
\end{equation}
noting that for $X=1$, there are 3 collisions but $N$ is precisely equal to 4 which precludes us from using the floor function. Substituting \eqref{eq:def_N} into \eqref{eq:def_N_tot} we find that the integer $ N_{\rm total} $ consists of $ d $ digits corresponding to first $ d $ digits of $ \pi $.

\newpage
\section*{Discussion and Conclusions}
\vspace{-12pt}
The framework that we have developed is analogous to Kepler's Second Law, which states \textit{``A line joining a planet and the Sun sweeps out equal areas in equal time''} \citep{1609anov.book.....K}. In fact, this problem has ties to Keplerian motion in another way, in that particle collisions may be treated as an approximation to the dynamics of a satellite performing a slingshot maneuver around a massive object.\footnote{\url{http://www.physics.usyd.edu.au/teach_res/mp/doc/mec_slingshot.htm}}

While the mass required to compute consecutive digits of $ \pi $ grows exponentially, rendering physical implementation of this method untenable, it is, nevertheless, possible to determine the first few digits of $\pi$ through physical experiments. The most famous example is \textit{Buffon's Needle} \citep{de1777essai} where a needle is dropped onto floorboards. The fraction of needles that lie across two boards is then dependent on the length of the needles, the distance between floorboards (which are both measurable quantities) and $\pi$. This method has been used to calculate $\pi$ to 5 decimal places using only three thousand needles \citep{lazzarini1901applicazione}. 

We note that while this experiment is entirely non-physical, qualitatively similar phenomena do occur in nature. Explosive astrophysical events such as solar flares \citep{1985ApJ...298..400E}, supernovae \citep{1992ApJ...399L..75R} produce observable outflows and shock waves. The primary mechanism by which these shock waves are driven is thought to be Fermi acceleration, where charged particles (analogous to the masses in our experiment) are repeatedly reflected by a magnetic mirror (analogous to the wall). This is also thought to be the origin of cosmic rays \citep{1987PhR...154....1B}.

We have found a solution to the unusual method of calculating $\pi$ proposed by \citet{galperin2003}. By considering constraints on the velocity phase-space of the system imposed by the conservation of energy and momentum, we take a geometric approach to show that if the ratio of masses of the two objects is of the form $10^{2(1-d)}$ then the number of times they collide is given by the first $d$ digits of $\pi$.  

\vspace{-12pt}
\section*{Acknowledgements}
\vspace{-12pt}
We thank A. Zic and M. Wheatland for useful discussions. We thank Grant Sanderson of \textit{3Blue1Brown} for bringing our attention to the topic.

MZR is supported by the Australian Research Council through grant DP160102932. DD is supported by an Australian Government Research Training Program Scholarship. This research has made use of NASA's Astrophysics Data System Bibliographic Services.

\textit{Software:} Matplotlib \citep{2007CSE.....9...90H}, Numpy \citep{2011CSE....13b..22V}

\vspace{-12pt}

\renewcommand{\refname}{{\large References\vspace{-18pt}}}
\bibliographystyle{apj_nofirst}
\footnotesize
\bibliography{bibliography}

\end{document}